\documentclass[sigconf]{acmart}

\setcopyright{none}
\renewcommand\footnotetextcopyrightpermission[1]{}

\usepackage[english]{babel}
\usepackage{xspace}
\usepackage{placeins}
\usepackage{mathtools}

\DeclarePairedDelimiter\floor{\lfloor}{\rfloor}

\newcommand{\xhdr}[1]{\vspace{.5\baselineskip}\noindent\textbf{#1}}

\newcommand{\coeq}{CO\textsubscript{2}eq\xspace}
\newcommand{\tone}{\textit{Discount}\xspace}
\newcommand{\ttwo}{\textit{Surcharge}\xspace}
\newcommand{\tthree}{\textit{Combined}\xspace}
\newcommand{\alpine}{C\textsubscript{1}\xspace}
\newcommand{\esplanade}{C\textsubscript{3}\xspace}
\newcommand{\piano}{C\textsubscript{2}\xspace}
\newcommand{\orni}{C\textsubscript{4}\xspace}
\newcommand{\numtransactions}{20,218\xspace}
\newcommand{\numtransactionsall}{40,087\xspace}
\begin{document}
\title{Carrot, stick, or both? Price incentives for sustainable food choice in competitive environments}

\author{Francesco Salvi}
\authornote{Work done at EPFL.}
\email{fsalvi@princeton.edu}
\affiliation{%
  \institution{Princeton University}
  \city{}
  \state{}
  \country{}
}

\author{Giuseppe Russo}
\email{giuseppe.russo@epfl.ch}
\affiliation{%
  \institution{EPFL}
  \city{}
  \state{}
  \country{}
}

\author{Adam Barla}
\email{adam.barla@epfl.ch}
\affiliation{%
  \institution{EPFL}
  \city{}
  \state{}
  \country{}
}

\author{Vincent Moreau}
\email{vincent.moreau@epfl.ch}
\affiliation{%
  \institution{EPFL}
  \city{}
  \state{}
  \country{}
}

\author{Robert West}
\email{robert.west@epfl.ch}
\affiliation{%
  \institution{EPFL}
  \city{}
  \state{}
  \country{}
}

\renewcommand{\shortauthors}{Salvi et al.}

\begin{abstract}
\textbf{Significance Statement.}
Pricing interventions are widely discussed as a lever for promoting sustainable food choices, yet most evidence comes from captive or highly constrained settings where consumers have limited alternatives. Our large-scale field experiment, conducted in a competitive real-world campus environment, reveals that pricing strategies can meaningfully reduce greenhouse gas emissions but may also trigger behavioral spillovers that undermine their environmental benefits. By directly comparing discounts, surcharges, and combined schemes, we show that balanced pricing can achieve climate impact without harming economic viability. These findings offer actionable insights for policymakers and institutions aiming to design effective and scalable sustainability interventions.

\vspace{0.5em}
\noindent\textbf{Abstract.}
Meat consumption is a major driver of global greenhouse gas emissions. 
While pricing interventions have shown potential to reduce meat intake, previous studies have focused on highly constrained environments with limited consumer choice.
Here, we present the first large-scale field experiment to evaluate multiple pricing interventions in a real-world, competitive setting.
Using a sequential crossover design with matched menus in a Swiss university campus, we systematically compared vegetarian-meal discounts (–2.5 CHF), meat surcharges (+2.5 CHF), and a combined scheme (–1.2 CHF/+1.2 CHF) across four campus cafeterias.
Only the surcharge and combined interventions led to significant increases in vegetarian meal uptake--by 26.4\% and 16.6\%, respectively--and reduced \coeq emissions per meal by 7.4\% and 11.3\%, respectively. 
The surcharge, while effective, triggered a 12.3\% drop in sales at intervention sites and a corresponding 14.9\% increase in non-treated locations, hence causing a spillover effect that completely offset environmental gains. 
In contrast, the combined approach achieved meaningful emission reductions without significant effects on overall sales or revenue, making it both effective and economically viable.
Notably, pricing interventions were equally effective for both vegetarian-leaning customers and habitual meat-eaters, stimulating change even within entrenched dietary habits.
Our results show that balanced pricing strategies can reduce the carbon footprint of realistic food environments, but require coordinated implementation to maximize climate benefits and avoid unintended spillover effects.
\end{abstract}
     
\keywords{Nutrition | Sustainability | GHG emissions | Climate change | Randomized Controlled Trials}

\maketitle
\balance
\section{Introduction}
The global food system is responsible for one-quarter to one-third of all anthropogenic greenhouse gas (GHG) emissions~\cite{Poore2018, fao2021, Crippa2021, IPCC2019}. In 2019 alone, according to estimates from the Food and Agriculture Organization of the United Nations, food and agriculture generated 17 billion tonnes of CO\textsubscript{2}-equivalent (\coeq), corresponding to about 31\% of global GHG emissions~\cite{fao2021}. 
Crucially, current food production and consumption trends are incompatible with international climate targets: to limit global warming to 1.5°C, the aspirational goal of the Paris Agreement~\cite{ParisAgreement2015}, overall emissions must fall by around 43-45\% by 2030 and ultimately reach net-zero by 2050. 
In contrast, food-related emissions have increased by 16\% since 1990, driven by growing demand and changing consumption patterns \cite{fao2021}.
This makes the food sector a critical target for climate action, with an urgent need for substantial interventions.

One of the most effective and immediate ways to address this challenge lies in dietary choices, as plant-based foods have a 10- to 50-fold lower carbon footprint per kilogram than animal products~\cite{Poore2018}. 
Consequently, even modest population-level shifts toward more vegetarian diets can yield significant reductions in emissions. 

Previous research has investigated a wide range of food-related interventions, including modifying option availability, such as increasing the proportion of vegetarian options~\cite{Garnett2019, Pechey2022} or introducing meat-free days~\cite{Russo2025, meat_free_movement, meatless_monday_origin, veggie_day_ghent}, as well as implementing informational cues like calorie counts, eco-labels or sustainability messaging~\cite{Clarke2025, Meier2022, CampbellArvai2024, Visschers2015, Ye2021, Thorndike2014}. Additional behavioral nudges include making plant-based meals the default option or improving their placement on menus~\cite{Boronowsky2022, CampbellArvai2024, Hnninghaus2025}, using appealing menu descriptions~\cite{Weijers2024, Gavrieli2022}, and leveraging social norms through peer influence~\cite{Sparkman2017}.

One particular type of intervention that has received growing attention is the use of price-based strategies~\cite{An2013, Andreyeva2022, Niebylski2015}. These approaches include subsidies for vegetarian options, taxes on high-carbon foods, or mixed pricing schemes that reward sustainable choices and penalize less sustainable ones, leveraging customers' well-established sensitivity to price variations~\cite{Garnett2021, Handziuk2023, Vellinga2022, Roy2021, Pizzo2024, Michels2008, Deliens2016, French1997}. 
Field experiments, often conducted in institutional settings such as universities and workplace cafeterias, have demonstrated the potential of such strategies as an effective lever for shifting food choices.
For example, a field experiment at the University of Cambridge combined an 8\% price increase on meat dishes with a 10\% discount on vegetarian menus, resulting in a 3.2\% increase in vegetarian sales but no significant reduction in meat purchases~\cite{Garnett2021}.
At HEC Paris Business School, a bonus-malus pricing policy directly linked dish prices to their carbon footprint, leading to a drop between 27\% and 42\% in the average footprint of purchased meals~\cite{Handziuk2023}.
Other research has shown that substantial discounts on vegetarian or healthy foods can significantly increase the sales of those items~\cite{French1997, Michels2008}, while moderate price increases on meat can reduce demand if paired with clear sustainability messaging~\cite{Vellinga2022, Pizzo2024}. 

Despite encouraging findings, research on price-based interventions remains limited in several important ways. 
Previous studies have focused on constrained settings such as single cafeterias~\cite{Garnett2019, Handziuk2023, Michels2008, Deliens2016}, online shops~\cite{Hoenink2020}, simulated environments~\cite{Vellinga2022}, or remote institutional venues~\cite{Pizzo2024, French1997}, where customers typically have few or no external alternatives and are effectively required to purchase within the study site. This captive setup may overstate the impact of price changes compared to real-world conditions, where customers can easily choose among many outlets operating in a competitive landscape. 
Moreover, much of the existing literature examines only one type of pricing intervention at a time and measures success mainly by the share of plant-based sales, without head-to-head comparisons between different strategies and less attention to actual emission reductions.

To address these gaps, we conducted a large-scale field experiment at EPFL, a major public research university in Switzerland. 
We evaluated the effectiveness of three distinct pricing strategies using a matched-menu design and a sequential crossover approach: (1) a 2.5 CHF discount on vegetarian meals, (2) a 2.5 CHF surcharge on meat-based meals, and (3) a hybrid scheme combining a 1.2 CHF discount on vegetarian meals with a 1.2 CHF surcharge on meat-based meals.
Our intervention was implemented in only four out of 12 campus dining facilities, representing approximately 50\% of total campus food sales during the intervention period. This setup intentionally allowed customers access to multiple alternative dining options, both within the university and in the surrounding urban area, hence mirroring the competitive conditions typical of real-world food environments. 
By tracking both purchasing behavior and associated GHG emissions, our study not only compares the effectiveness of multiple pricing strategies but also provides direct estimates of their potential climate impact in a realistic, competitive setting.

Our results reveal marked differences in the effectiveness and trade-offs associated with each pricing strategy. 
Discounts on vegetarian meals did not produce a significant change in the proportion of vegetarian meals sold ($p=0.58$) or in \coeq emissions ($p=0.56$). 
In addition, while this approach attracted additional customers and modestly increased total sales, the corresponding loss in revenue outweighed these gains, rendering the intervention economically unsustainable for food providers in the long term.
In contrast, surcharges on meat-based meals proved highly effective at driving dietary change. This intervention produced a substantial 26.4\% increase (95\% CI [+19.0\%, +33.9\%], $p < 0.0001$) in the uptake of vegetarian meals and led to notable reductions in \coeq emissions, with a 7.4\% decrease per meal (95\% CI [-14.8\%, -0.01\%], $p = 0.049$). 
However, these environmental gains came at the cost of considerable customer churn, as many customers shifted their purchases to other campus cafeterias not affected by the price increase. This led to a 12.3\% reduction (95\% CI [-17.9\%, -6.6\%], $p=0.0001$) in sales at intervention sites and a comparable 14.9\% increase (95\% CI [+1.5\%, +28.3\%], $p=0.03$) in non-treated cafeterias. In those non-treated sites, \coeq emissions also rose by 12.1\% (95\% CI [+8.4\%, +15.9\%], $p < 0.0001$), largely neutralizing the positive environmental impact of the surcharges.
The combined strategy, which featured moderate price adjustments for both vegetarian and meat-based meals, emerged as the most balanced and promising intervention. 
The combined approach yielded a 16.6\% increase (95\% CI [+9.5\%. +23.7\%],  $p < 0.0001$) in vegetarian meal sales and an 11.3\% reduction (95\% CI [-18.2\%, -4.4\%] $p=0.001$) in \coeq emissions per meal.
Importantly, it did so without triggering significant losses in either sales ($p=0.28$) or revenue ($p=0.19$), demonstrating both economic and practical sustainability.

Our results suggest that effective policies should simultaneously incentivize vegetarian choices and discourage meat consumption, offering key insights for decision-makers and policymakers.

\section{Methods}
\xhdr{Experimental design.}
\begin{figure*}[th]
    \centering
    \includegraphics[width=.85\linewidth]{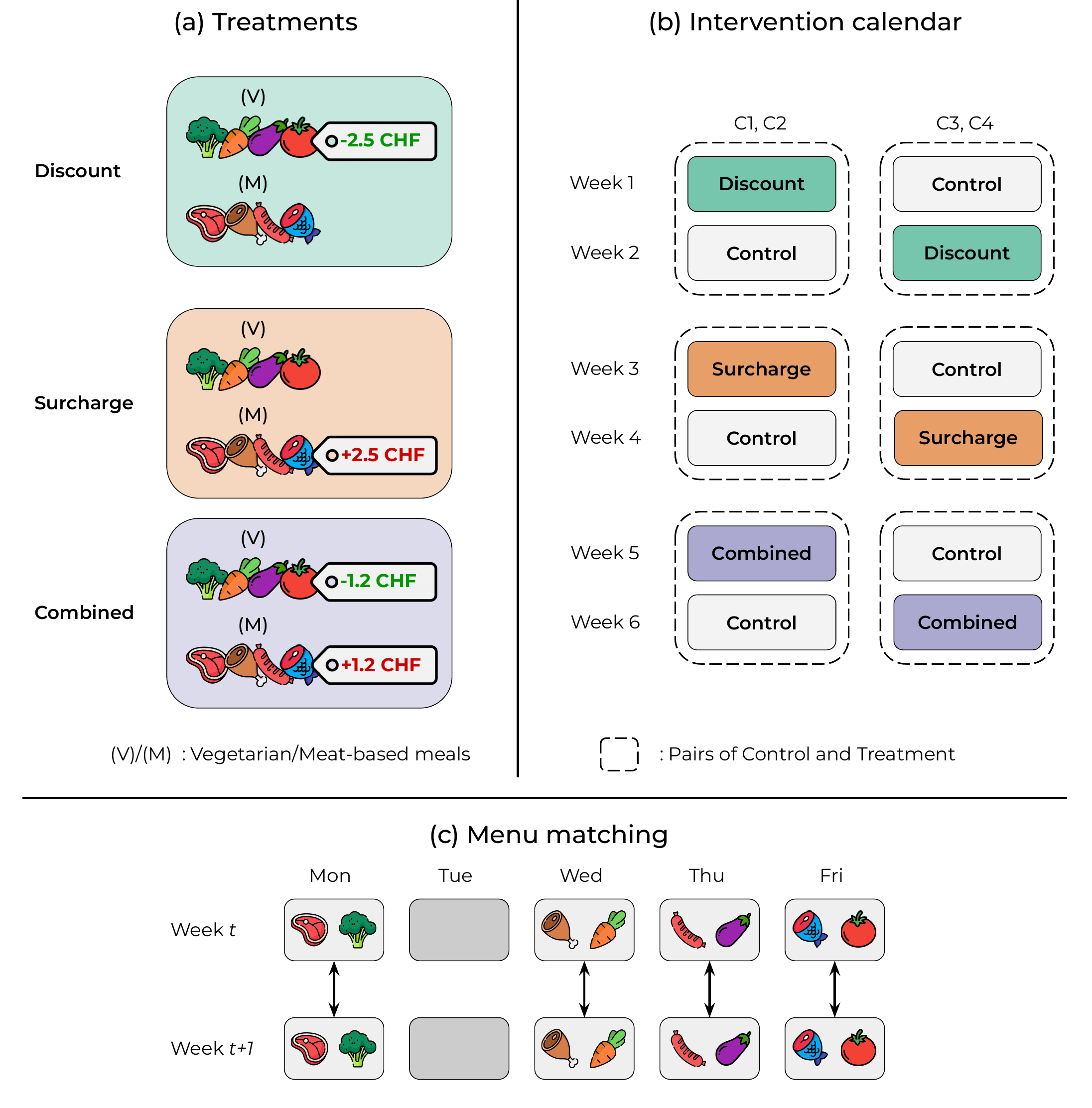}
    \Description{}{}
    \caption{
    \textbf{Experimental design.}
    (a) We consider three pricing strategies: (\tone) a 2.5 CHF price reduction on vegetarian meals, (\ttwo) a 2.5 CHF price increase on meat-based meals, and (\tthree) a hybrid scheme with a 1.2 CHF reduction for vegetarian meals and a 1.2 CHF increase for meat-based meals. 
    (b) Weekly assignment of treatment and control conditions for each cafeteria (\alpine, \piano; \esplanade, \orni) over the six-week study period. Each cafeteria alternates between control and treatment weeks within each phase, following a crossover design and serving as its own control. 
    (c) Schematic representation of menu standardization across weeks. Menus are matched on each day of the week (except Tuesday, due to campus-wide vegetarian policy), independently for each cafeteria, ensuring comparability between control and treatment periods.
}
    \label{fig:design}
\end{figure*}
Our experiment took place at EPFL, a large university in Switzerland with an overall population of more than 18,000 individuals (see \autoref{app:epfl} for additional information about EPFL’s population and its food ecosystem). Specifically, we selected four out of the 12 campus cafeterias---named in the following \alpine, \piano, \esplanade, and \orni---which together accounted for approximately 50\% of total campus food sales during the study period. 

Our experimental design is summarized schematically in \autoref{fig:design}.
The experiment spanned six weeks, from October 28 to December~6, 2024, coinciding with standard lecture weeks on the academic calendar. This timing intentionally excluded holiday breaks and exam sessions to ensure consistent levels of on-campus activity.

We divided the study period into three consecutive phases, each corresponding to a distinct pricing intervention (see \hyperref[fig:design]{Figure 1a}):
\begin{enumerate}
    \item \textbf{\tone} (Weeks 1-2): A -2.5 CHF price reduction for vegetarian meals.
    \item \textbf{\ttwo} (Weeks 3-4): A +2.5 CHF price increase for meat-based meals\footnote{For the purpose of this paper, ``meat-based'' also includes fish.}. 
    \item \textbf{\tthree} (Weeks 5-6): A hybrid strategy combining a -1.2 CHF price reduction for vegetarian meals with a +1.2 CHF price increase for meat-based meals.
\end{enumerate}
We note that, for the \tthree strategy, the value of 1.2 CHF was chosen instead of the more natural midpoint of 1.25 CHF, since all prices at EPFL cafeterias are rounded to the first decimal place.
Considering that normal prices range from 6.1 to 16.3 CHF, with discounts available for students, the price variations in the \tone and \ttwo interventions range between 15\% and 40\% of the original price, and the \tthree variation between 7\% and 20\% of the original price.

Within each two-week phase, each cafeteria alternated between a one-week control period and a one-week treatment period (see \hyperref[fig:design]{Figure 1b}), following a crossover design where every cafeteria served as its own control. For example, during the first phase, \alpine and \piano were assigned to the \tone intervention during Week 1 and acted as controls in Week 2. Vice versa, \esplanade and \orni served as controls in Week 1 and were assigned to the \tone intervention in Week 2. This pattern was applied consistently across all three phases, consistently pairing \alpine with \piano and \esplanade with \orni.

\begin{sloppypar}
To isolate the effect of our interventions, we collaborated closely with caterers to standardize menu offerings in a matched design (see \hyperref[fig:design]{Figure 1c}). 
For each cafeteria, menus were held constant on the same weekday across the two weeks of each phase (e.g., Monday of Week 1 and Monday of Week 2), ensuring that differences between control and treatment conditions reflected the effect of the price intervention rather than variation in meal offerings.
Formally, for all cafeterias $c \in  C= \{$\alpine, \piano, \esplanade, \orni\unskip$\}$, and for each day $d$ in weeks $t \in \{1, 3, 5\}$, the menus satisfied $M_{c, t, d} = M_{c, t+1, d}$. 
In general, given $c \neq c' \in C$, $M_{c, t, d} \neq M_{c', t, d}$. That is, the matching process happens independently for each cafeteria, with a different set of menus.
The experiment excluded all Tuesdays, because of a pre-existing, campus-wide policy enforcing only vegetarian dishes to be sold in all cafeterias on Tuesdays~\cite{Russo2025}. On all other weekdays, each cafeteria consistently offered at least one vegetarian and one meat-based option.
\end{sloppypar}

Campus members were notified in advance of our study through a newsletter from EPFL's Food Commission and a web page detailing our initiative. Our protocol was approved by EPFL's Human Research Ethics Committee (HREC 074-2023).

\xhdr{Data.}
We leveraged EPFL's advanced food data monitoring infrastructure to assemble a comprehensive, anonymized dataset of food purchases made on campus during the experimental period. This dataset covers all permanent food outlets, including cafeterias, food trucks, and vending machines.
Each transaction record contains the date, time, and location of the purchase, the specific items bought, their prices, and a unique anonymized identifier for each customer, enabling us to track individual purchasing patterns over time. 
For each anonymized identifier, we have access to broad demographic attributes of the respective customer, including their gender, age, and campus status (student, PhD student, or staff).
Transactions are also enriched with detailed menu information: meal descriptions, ingredient lists with quantities, and precise GHG emission estimates. 
GHG estimates were provided by a specialized Swiss company and are based on actual ingredients purchased by campus caterers, accounting for agricultural production, seasonality, transportation, and packaging~\cite{beelong}. This results in a high level of precision that is unmatched in previous research.
Meals were classified as vegetarian or meat-based according to their ingredient lists. 
Our final dataset includes \numtransactionsall transactions recorded during the six-week intervention period.

\xhdr{Statistical analysis.}
To evaluate the effects of the interventions, we analyzed two categories of outcomes: environmental and economic.

Environmental outcomes capture changes in meal choices and their associated \coeq emissions.
We modeled two dependent variables: the probability of purchasing a vegetarian meal and the per-transaction emissions, with treatment indicators as predictors.
We estimated these effects using two-way fixed-effects regressions~\cite{Wooldridge2010, Greene2012, russo2023spillover} (full model specifications in \autoref{app:model}), including fixed effects for both customer and date to account for repeated observations and unobserved heterogeneity caused by day-specific factors such as weather or campus events. 

Economic outcomes capture changes in cafeteria performance, measured by total daily sales and gross revenue. 
To adjust for baseline differences in sales volume across cafeterias and weekdays, we normalized each observation by its value during the corresponding control period. 
We then modeled normalized sales and gross revenue as dependent variables, again with treatment indicators as predictors, and we estimated treatment effects using fixed-effects regressions (full model specifications in \autoref{app:model}).
Here, only date fixed effects were included, as the normalization procedure already absorbed cafeteria-level differences. 

\section{Results}
\begin{figure*}[th]
    \centering
    \includegraphics[width=.85\linewidth]{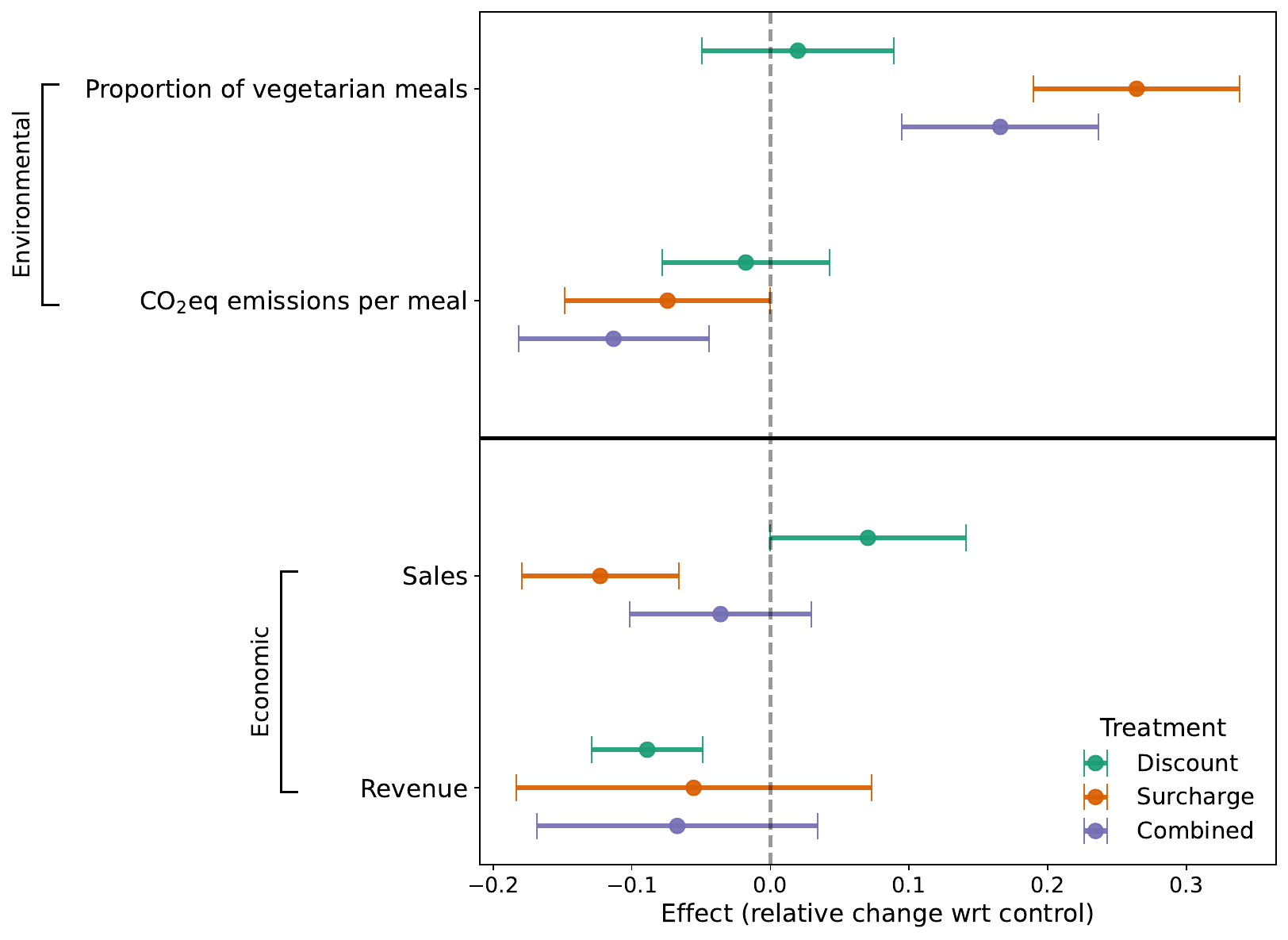}
    \Description{}{}
    \caption{
    \textbf{Regression results.}
    Environmental outcomes (proportion of vegetarian meals and \coeq emissions per meal) were analyzed using two-way fixed-effects regressions on customers and dates. 
    Economic outcomes (sales and gross revenue) were aggregated and standardized at the daily and cafeteria level, and intervention effects were estimated using fixed-effects regressions on dates.
    All results are expressed as relative changes compared to the control period. Error bars indicate 95\% confidence intervals. Full model specifications in \autoref{app:model}.
}
    \label{fig:effects}
\end{figure*}

\xhdr{Environmental outcomes.}
We report in \autoref{fig:effects}~(top) the environmental effects of our interventions.
During the experimental period, the proportion of vegetarian meals sold increased from a baseline of 0.410 in the control period to 0.418 under the \tone intervention (+2.0\%, 95\% CI [-4.9\%, +8.9\%], $p=0.58$), 0.520 under the \ttwo intervention (+26.4\%, 95\% CI [+19.0\%, +33.9\%], $p < 0.0001$), and 0.478 under the \tthree intervention (+16.6\%, 95\% CI [+9.5\%. +23.7\%],  $p < 0.0001$).
A similar pattern was observed for GHG emissions. Average \coeq emissions per meal declined from 1.64 kg during the control period to 1.61 kg during the \tone intervention (-1.8\%, 95\% CI [-7.8\%, +4.2\%, $p=0.56$]), to 1.52 kg under the \ttwo intervention (-7.4\%, 95\% CI [-14.8\%, -0.01\%], $p = 0.049$), and to 1.46 kg during the \tthree intervention (-11.3\% 95\% CI [-18.2\%, -4.4\%] $p=0.001$).
These results highlight the heterogeneity in the effectiveness of different pricing strategies. The \tone intervention did not yield statistically significant changes in either vegetarian meal uptake or average \coeq emissions. In contrast, both \ttwo and \tthree interventions led to significant reductions in \coeq emissions, primarily driven by increased selection of vegetarian meals.

\xhdr{Economic outcomes.}
To assess the practical and economic sustainability of our interventions, we also analyzed daily sales and gross revenue during the experimental period.
As shown in \autoref{fig:effects}~(bottom), the \tone intervention produced a 7.0\% increase in sales compared to the control period (95\% CI [-0.05\%, +14.1\%], $p=0.5$, but this was accompanied by an 8.9\% decrease in revenue (95\% CI [-12.9\%, -4.9\%], $p <0.0001$). 
This suggests that although more meals were sold, possibly by attracting new customers or increasing the frequency of purchases, the increase in sales was not enough to compensate for the revenue loss due to the discounted prices of vegetarian options.
Conversely, the \ttwo intervention resulted in a substantial 12.3\% reduction in sales (95\% CI [-17.9\%, -6.6\%], $p=0.0001$), indicating a notable customer churn. However, the revenue drop was not statistically significant (-5.5\%, 95\% CI [-18.3\%, +7.2\%], $p=0.39$), likely because the increased prices of meat-based options partially offset the loss in volume.
Finally, the \tthree intervention produced a modest, non-significant decrease in sales (-3.6\%, 95\% CI [-10.1\%, +3.0\%], $p=0.28$) and a non-significant decline in revenue (-6.7\%, 95\% CI [-16.8\%, +3.4\%], $p=0.19$). This suggests a balance between the previous two mechanisms: the reduction in meat-based meal purchases and increase in vegetarian sales, coupled with moderate pricing changes, resulted in a significant loss in neither customers nor gross revenue.

\xhdr{Spillover effects.}
To further investigate the practical repercussions of our interventions, we extended our analysis to the eight EPFL cafeterias that were not included in the experiment. These venues act as direct commercial competitors, since campus members have access to all 12 cafeterias and often select the one that best meets their preferences and convenience on any given day. 
To assess potential spillover effects, we replicated our analysis for both environmental and economic outcomes in these cafeterias, using a six-week pre-experimental period, from September~9 to October~19, 2024, as the control baseline (see \autoref{app:model} for full technical details). It is worth noting that, in this case, only daily aggregated outcomes (sales and revenue) allow for robust interpretation, as the environmental analysis is limited by the lack of a matched-menu design (cf. \autoref{fig:design}). Thus, environmental findings on spillover effects should be regarded as suggestive rather than conclusive.

Our results show no significant changes in sales or gross revenue in the non-treated cafeterias for both the \tone (+1.1\% sales, 95\%  CI [-16.1\%, +18.2\%], $p=0.90$; +1.8\% revenue, 95\% CI [-16.5\%, +20.1\%], $p=0.85$) and the \tthree interventions (+2.8\% sales, 95\% CI [-10.9\%, +16.4\%], $p=0.69$; +2.3\% revenue, 95\% CI [-9.3\%, +14.0\%], $p=0.70$). These findings are in line with results from the intervention cafeterias, where neither intervention led to significant changes in overall sales or revenue.

\begin{sloppypar}
By contrast, the \ttwo intervention produced a marked spillover: sales in the eight non-treated cafeterias increased by 14.9\% (95\% CI [+1.5\%, +28.3\%], $p=0.03$), and revenue rose by 14.8\% (95\% CI [+0.3\%, +29.4\%], $p=0.04$). 
This sharp increase mirrors the significant decline in sales observed in the four intervention cafeterias, suggesting that many customers simply shifted their purchases to other locations when faced with higher prices for meat-based meals. This customer migration has important implications not only for the economic sustainability of the \ttwo intervention, but also for its overall effectiveness in reducing environmental impact. 
Indeed, during the \ttwo period, the proportion of vegetarian meals sold in the non-treated cafeterias dropped by 8.6\% (95\% CI [-10.9\%, -6.2\%], $p < 0.0001$), while average per-meal \coeq emissions increased by 12.1\% (95\% CI [+8.4\%, +15.9\%], $p < 0.0001$), virtually offsetting the carbon savings achieved in the intervention sites.
\end{sloppypar}

\xhdr{Sociodemographic patterns.}
\begin{figure*}[p]
    \centering
    \includegraphics[width=.752\linewidth]{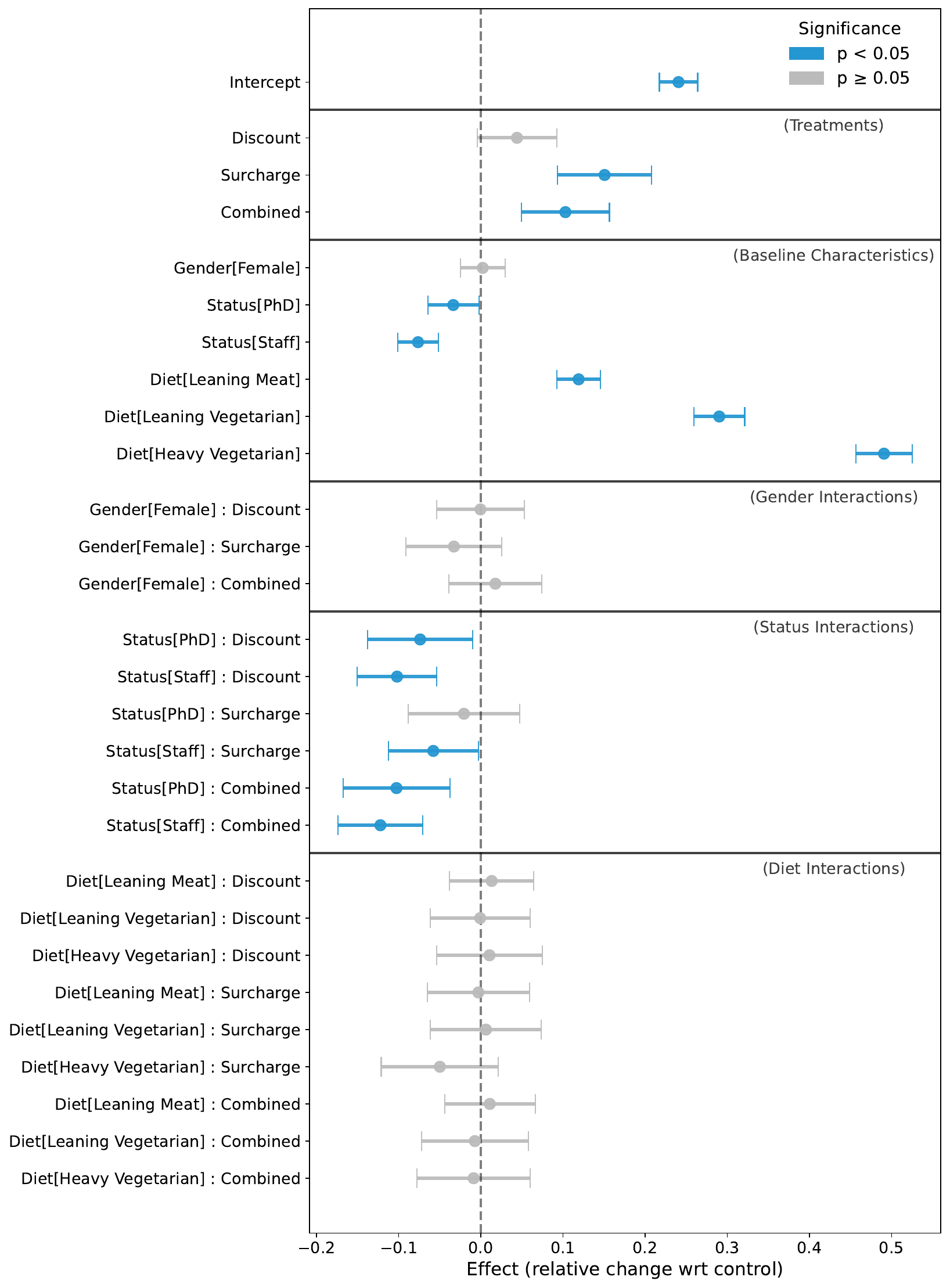}
    \Description{}{}
    \caption{
    \textbf{Regression results---proportion of vegetarian meals as a function of individual profiles.}
    The proportion of vegetarian meals was modeled using a fixed-effects regression with date fixed effects and interaction terms for sociodemographic factors (gender, campus status, and dietary history).
    The \textit{Diet} attribute categorizes individuals based on their percentage of vegetarian meals purchased during a six-week pre-intervention period, including only individuals with at least five transactions.
    Coefficients are shown relative to the reference category: male students with a heavy meat diet. 
    Error bars indicate 95\% confidence intervals.
    Full model specifications in \autoref{app:model}.    
}
    \label{fig:individual_effects}
\end{figure*}
Leveraging the rich detail in the data collected through EPFL's infrastructure, we explored whether the effect of our pricing interventions varied systematically by individual characteristics. Specifically, we examined three key attributes: gender (restricted to male or female for this analysis), on-campus status (student, PhD student, or staff), and previous dietary preferences.
To quantify prior dietary habits, we analyzed each individual’s food purchases over the same six-week pre-experimental period used for spillover analyses (September 9 to October 19, 2024). Individuals were categorized into a categorical attribute (\textit{Diet}) with four levels based on the percentage of vegetarian meals they had purchased: \textit{Heavy Meat} (0–25\% vegetarian purchases), \textit{Leaning Meat} (25–50\%), \textit{Leaning Vegetarian} (50–75\%), or \textit{Heavy Vegetarian} (75–100\%). For this analysis, we only included individuals with at least five recorded transactions during this period.
We then estimated the probability of selecting a vegetarian meal using the same fixed-effects regression described previously and specified in \autoref{app:model}, but replacing customer-specific fixed effects with controls for the sociodemographic attributes and their interactions with the treatments.
The resulting estimates, reported in \autoref{fig:individual_effects}, represent changes in the share of vegetarian meals purchased relative to the reference condition: a male student with a heavy meat diet during the control period. For instance, an effect size of +0.10 indicates a 10 percentage point increase over the intercept in the share of vegetarian meals. 
Overall, we found that the treatment effects for each pricing intervention when accounting for these sociodemographic factors remained robust and consistent with those reported in \autoref{fig:effects}.
In particular, the \tone intervention produced a small, non-significant increase in the share of vegetarian meals ($p=0.08$), while the \ttwo and \tthree interventions led to substantial increases, raising the share of vegetarian meals by 15.1 and 10.3 percentage points, respectively ($p < 0.0001$ and $p=0.0002$).
No meaningful differences were observed by gender, either in overall vegetarian meal selection ($p=0.87$) or in response to the interventions (all interaction $p$-values between 0.27 and 0.98).
In contrast, significant differences emerged by campus status. During the control period, PhD students and staff chose vegetarian meals less frequently than students (by 3.2 and 7.7 percentage points, respectively; $p=0.03$ and $p<0.0001$), suggesting a greater inclination toward meat-heavy diets among more senior campus members. In addition, the interventions were systematically less effective for staff members: compared to students, increases in vegetarian meal selection were smaller among staff under all three interventions (by 10.0, 5.8, and 12.2 percentage points, respectively; $p < 0.0001$, $p=0.03$, and $p < 0.0001$). PhD students showed mixed results: less affected by the \tone (by 7.4 percentage points, $p=0.02$) and \tthree (by 10.0  percentage points, $p=0.002$) interventions than students, but still responding to the \ttwo intervention with no significant difference relative to students ($p=0.55$). 
These results indicate that price sensitivity is closely tied to campus status and, by extension, likely to income or purchasing power. Staff members, who tend to have higher and more stable incomes, were less influenced by changes in meal prices, while students remained highly responsive.
Finally, as expected, we observe that individuals with a higher past tendency to consume vegetarian meals were more likely to choose vegetarian options overall. 
Crucially, there was no evidence that dietary habits had any effect on the effectiveness of the interventions, with all groups responding equally to price changes (all interaction $p$-values between 0.17 and 0.98). This suggests that pricing interventions can effectively shift meal choices even among those who typically prefer meat-heavy diets, proving to be a powerful nudge among habitual meat eaters.

\section{Discussion}
In this study, we conducted one of the largest field experiments to date on price-based interventions for promoting sustainable food choices in a real-world, competitive setting. 
Leveraging a sequential crossover design with matched menus, we systematically compared the effect of vegetarian meal discounts (-2.5 CHF), surcharges on meat-based options (+2.5 CHF), and a combined strategy (-1.2 CHF and +1.2 CHF), providing the first comprehensive head-to-head comparison of heterogeneous pricing strategies. 
Unlike previous literature, our intervention was embedded in a complex, multi-venue food environment, where customers were free to ``vote with their feet'' among multiple on- and off-campus alternatives. This unique setup allowed us to assess both the direct impact of pricing interventions on \coeq emissions and their indirect effects, including customer spillover, economic sustainability, and heterogeneity across demographic groups, in conditions that closely mirror the practical realities faced by institutional food providers. 

Our findings indicate that surcharges on meat-based meals, either alone or in combination with moderate discounts on vegetarian options, can effectively shift dietary choices and reduce greenhouse gas emissions. However, each pricing strategy entailed distinct economic trade-offs: pure discounts risk eroding revenue, while pure surcharges risk losing customers and generating spillover effects that offset environmental gains. The combined approach, instead, emerged as the most promising intervention, offering a balanced compromise that can achieve environmental objectives without significant adverse effects on sales or gross revenue. Notably, this \tthree intervention yielded a substantial 16.6\% increase (95\% CI [+9.5\%. +23.7\%],  $p < 0.0001$) in vegetarian meal sales and an 11.3\% reduction (95\% CI [-18.2\%, -4.4\%] $p=0.001$) in \coeq emissions per meal.
Our analysis of sociodemographic and behavioral patterns further revealed that price sensitivity varied strongly by campus status: senior members, who tend to have higher incomes, were less responsive to pricing changes than students. Notably, the interventions were similarly effective across individuals with predominantly meat-heavy diets, indicating that even habitual meat-eaters can be nudged toward more sustainable choices in the presence of appropriate incentives.

\xhdr{Societal implications.}
These findings carry important implications for both institutional decision-makers and policymakers aiming to reduce the carbon footprint of food environments. 
Our results suggest that carefully calibrated pricing interventions can serve as powerful levers for advancing climate goals, without sacrificing economic viability or customer satisfaction. In particular, the effectiveness of a balanced, combined approach highlights the value of nuanced policies that simultaneously incentivize vegetarian choices while discouraging meat-based ones.
To illustrate the practical impact, if our \tthree intervention had been applied to all cafeteria sales at EPFL in 2024, it would have reduced annual \coeq emissions by approximately 335 metric tons, corresponding to about 13\% of the entire carbon footprint of food estimated by EPFL for 2024 (2,619 metric tons)~\cite{epflDurabilite}. This saving is equivalent to the emissions generated by powering 70 U.S. homes for a year~\cite{EIA2023}, or by driving 1.37 million kilometers in an average gasoline car~\cite{EPA2024b}. Alternatively, it is equivalent to the average annual carbon sequestered by 5,542 urban tree seedlings each year over their first 10 years of growth~\cite{McPherson2016, US1998, krsteski2025valid}, or by 136 hectares of forest in one year~\cite{Smith2007, EPA2024}. 
Such tangible reductions show how institution-level interventions, when scaled, can make a meaningful contribution to climate mitigation efforts.
More broadly, our study underscores that even entrenched dietary habits, including those of habitual meat-eaters, can be shifted by targeted economic incentives. 
However, the pronounced spillover effects observed with simple meat surcharges highlight the importance of considering the broader competitive landscape: interventions implemented in isolation may simply redistribute unsustainable behaviors rather than reducing them. 
To maximize both environmental and economic outcomes, coordinated or campus-wide strategies, potentially in combination with other behavioral nudges or educational initiatives, are likely to be most effective. 
Our work demonstrates that effective climate action in the food sector requires not only ambitious goals but also a keen attention to behavioral realities and local context.

\xhdr{Limitations and future work.}
While our study offers robust insights into the real-world impact of pricing interventions, it nonetheless has limitations. 
First, although we employed a rigorous matched-menu design to ensure comparability between intervention and control periods, perfect menu matching was not always feasible in practice. Unforeseen operational constraints occasionally led caterers to introduce slight variations in menu offerings across corresponding days. While these deviations were infrequent and typically minor, they may have introduced some uncontrolled variation in meal choice and composition. 
Additionally, our experiment was conducted over a relatively short time frame (six weeks) and within a single institutional context, which may limit the generalizability of the findings to other settings, longer-term interventions, or different population groups. 

Future work could extend the duration of interventions to assess the long-term sustainability and potential for habituation or adaptation in dietary choices. 
Replicating the experiment across different institutional settings, such as schools, hospitals, or workplace canteens, would also offer valuable insights into the generalizability of our results and the influence of context-specific factors.

\begin{acks}
This research is supported financially by a grant from the EPFL Solutions4Sustainability initiative and was approved by the EPFL Human Research Ethics Committee (HREC 074-2023). The authors are grateful to the management of EPFL and, in particular, the entire team at the Food \& Beverages unit, in charge of all cafeterias on campus, and at the Identity and Access Management unit, who helped us gather purchasing data. We thank Mika Senghaas for his his preliminary work analyzing the data and Kristina Gligorić for helpful discussions. We would also like to thank all the caterers and those responsible for feeding EPFL staff and students every day.
\end{acks}

\FloatBarrier
\bibliographystyle{unsrt}
\bibliography{biblio}

\appendix
\section{Campus details}\label{app:epfl}
\begin{figure}[ht]
    \centering
    \includegraphics[width=\linewidth]{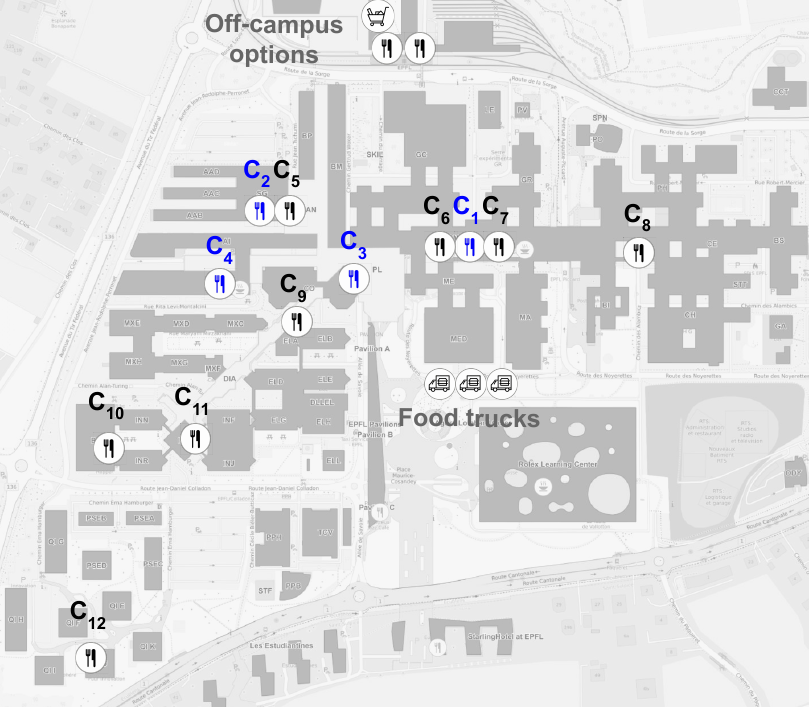}
    \Description{}{}
    \caption{
    \textbf{Map of food options on EPFL's campus}. 
    The campus features 12 cafeterias, distributed across the entire area. Additionally, customers have access to several food trucks and off-campus options, including a grocery store, a fast-food restaurant, and an Italian restaurant. The four cafeterias where the experiment took place are coloured in blue.
}
    \label{fig:map}
\end{figure}
As of 2024, EPFL (\textit{École Polytechnique Fédérale de Lausanne}) is home to a diverse community, consisting of 14,012 students and 6,477 employees, including professors, teachers, and administrative, technical, and scientific staff~\cite{epflfigures}. Representing over 130 nationalities, the population includes a 37\% female ratio, reflecting the university’s international and inclusive character.

To meet the needs of its large population, EPFL’s food ecosystem features 12 main cafeterias distributed across the central and peripheral areas of campus, as well as a number of food trucks that further diversify the food offering. The four cafeterias selected for our study collectively accounted for \numtransactions out of \numtransactionsall total cafeteria transactions (approximately 50\%) during the experimental period. In addition to these on-campus options, a fast-food restaurant and a grocery store are located within a short distance from the main campus buildings---between 100 and 500 meters away depending on the position on campus, which corresponds to a maximum walk of about seven minutes. This makes off-campus dining both convenient and easily accessible. Throughout our intervention, students and staff therefore had a wide array of choices, both on and off campus, ensuring that participation in the studied cafeterias was entirely voluntary and subject to normal competitive pressures.
The spatial arrangement of all food outlets, including cafeterias, food trucks, and nearby off-campus options, is visualized in \autoref{fig:map}.


\section{Model details}\label{app:model}
\xhdr{Environmental outcomes}
For the analysis of individual food choices and emissions, we used the complete transaction-level dataset ($N=\numtransactions$), analyzing both the probability of selecting a vegetarian meal (binary indicator) and the per-transaction \coeq emissions (continuous variable). 
For each outcome, we used the following two-way fixed effects regression model:
\begin{equation}\label{eq:transactions_model}
    Y_{i, c, d} = \alpha + \sum_{t \in T} \beta_t \text{Treatment}_{t, c, d} + \gamma_c + \delta_d + \varepsilon_{i,c,d}
\end{equation}
\begin{sloppypar}
where $Y_{i, c, d}$ is the outcome (vegetarian indicator or \coeq emissions) for transaction $i$, customer $c$ on date $d$; $\text{Treatment}_{t, c, d}$ are indicator variables for the three pricing interventions ($T = \{\text{\tone, \ttwo, \tthree\}}$, with the control period as the reference category; and $\varepsilon_{i,c,d}$ is the idiosyncratic error term.
$\gamma_c$ denotes the customer fixed effect, which controls for all unobserved, time-invariant characteristics of each customer, such as dietary preferences, daily schedules, and demographic attributes. Similarly, $\delta_d$ denotes the date fixed effect, which controls for all shocks or variations affecting all customers on a given day (e.g., menu offering, weather, or campus events). Standard errors were clustered by customer to account for possible residual correlation, ensuring more accurate inference even in the presence of repeated measurements~\cite{Wooldridge2010, Greene2012, russo2025pluralistic, russo2023understanding}.
\end{sloppypar}
When modeling sociodemographic patterns (see \autoref{fig:individual_effects}), we replaced $\gamma_c$ with individual controls, including also their interactions with $\text{Treatment}_{t, c, d}$.

\xhdr{Economic outcomes}
To analyze total sales and gross revenue, we aggregated the data at the level of each cafeteria and date ($N=96$). To account for baseline differences in sales volumes, prices, and customer flow between cafeterias and across weekdays, we normalized each observation using the following formula:
\begin{equation}\label{eq:aggregate_standardization}
    \tilde{Y}_{l, w, d} = \frac{Y_{l, w, d}}{Y_{l, \text{control}_w, d}}
\end{equation}
where $Y_{l, w, d}$ is the outcome (total sales or revenue) at location $l$, in week $w$, on weekday $d$, and ${Y}_{l, \text{control}_w, d}$ is the outcome at location $l$ on weekday $d$ during the control period corresponding to week $w$, depending on its experimental phase. Formally (cf. \autoref{fig:design}), 

$$
\text{control}_w =
\begin{cases}
    \floor*{(w+1) / 2}, & \text{if } l \in \text{\esplanade, \orni} \\
    \floor*{(w+1) / 2} + 1, & \text{if } l \in \text{\alpine, \piano}
\end{cases}
$$

This normalization expresses sales (or revenue) as a proportion of the location's own baseline for each weekday, so that an effect size of 1.10, for example, indicates a 10\% increase relative to the control week for that cafeteria and weekday. 

We then estimated treatment effects using the following fixed effects model:
\begin{equation}\label{eq:aggregate_model}
    \tilde{Y}_{l, w, d} = \alpha + \sum_{t \in T} \beta_t \text{Treatment}_{t, l, w} + \delta_d + \varepsilon_{l, w,d}
\end{equation}
where again $\text{Treatment}_{t, l, w}$ are indicator variables for the three pricing interventions, $\varepsilon_{l, w,d}$ is the residual error, and $\delta_d$ denotes the date fixed effect. Standard errors were clustered by cafeteria to account for potential correlations within each location over time.

\xhdr{Spillover analysis.}
To analyze potential spillover effects~\cite{russo2023spillover, russo2025does, russo2024shock}, we extended our analysis to the eight cafeterias that were not included in the experiment. 

To get a reference period for those cafeterias, we used as control a six-week pre-experimental period running from September 9 to October 19, 2024. Since no interventions ever ran within those cafeterias, we considered as treatments the entire two-week periods where interventions were active in at least one intervention cafeteria. In other words, for non-treated cafeterias the \tone period corresponds to Weeks 1-2, the \ttwo period to Weeks 3-4, and the \tthree period to Weeks 5-6 (cf. \autoref{fig:design}).

\begin{sloppypar}
We repeated the environmental transaction-level analysis in \eqref{eq:transactions_model}, with $\text{Treatment}_{t, c, d}$ defined as previously mentioned. For the aggregate economic analysis, we normalized outcomes by dividing observations by their corresponding value in the pre-experimental control period. This corresponds to the normalization formalized in \eqref{eq:aggregate_standardization}, with $\text{control}_w = w - 7$. Then, we repeated the analysis using the regression in \eqref{eq:aggregate_model}.
\end{sloppypar}

\end{document}